\documentclass[review]{elsarticle}

\usepackage{placeins}
\usepackage{rotating}
\usepackage{setspace}
\usepackage[utf8]{inputenc}
\usepackage[T1]{fontenc}
\usepackage{array}
\usepackage{booktabs}
\usepackage{tabulary}
\usepackage{longtable}
\usepackage{amsmath,amssymb}
\usepackage{graphicx}
\usepackage{caption}
\usepackage{subcaption}
\usepackage{url}
\usepackage{algorithm}
\usepackage{algpseudocode}
\usepackage{comment}

\renewcommand{\algorithmiccomment}[1]{\bgroup\hfill $\triangleright$~\parbox[t]{0.42\textwidth}{\scriptsize\begin{spacing}{1}#1\end{spacing}}\egroup}
\DeclareMathOperator{\bsize}{\mathsf{bsize}}
\DeclareMathOperator{\states}{\mathsf{vstates}}
\DeclareMathOperator{\dist}{\mathsf{dist}}
\DeclareMathOperator{\fitness}{\mathsf{fitness}}
\DeclareMathOperator{\maxvisits}{\mathsf{maxvisits}}
\DeclareMathOperator{\ssize}{\mathsf{ssize}}
\DeclareMathOperator{\knn}{k\mathsf{nn}}
\DeclareMathOperator{\N}{\mathcal{N}}
\DeclareMathOperator{\rand}{\mathsf{rand}}
\DeclareMathOperator{\recall}{\mathsf{recall}}
\DeclareMathOperator{\scale}{\mathsf{scale}}
\DeclareMathOperator{\mutate}{\mathsf{mutate}}
\DeclareMathOperator{\crossover}{\mathsf{crossover}}
\DeclareMathOperator{\ParetoRecall}{\mathsf{Pareto-recall}}
\DeclareMathOperator{\ParetoRadius}{\mathsf{Pareto-radius}}
\DeclareMathOperator{\MinRecall}{\mathsf{Min-recall}}

\begin{document}
\begin{frontmatter}

\title{Similarity search on neighbor's graphs with automatic Pareto optimal performance and minimum expected quality setups based on hyper-parameter optimization}

\author[conacyt,infotec]{Eric S. Tellez}
\ead{eric.tellez@infotec.mx}

\author[conacyt,centrogeo]{Guillermo Ruiz\corref{mycorrespondingauthor}}
\cortext[mycorrespondingauthor]{Corresponding author}
\ead{lgruiz@centrogeo.edu.mx}

\address[conacyt]{Consejo Nacional de Ciencia y Tecnología - Conacyt (México)}
\address[infotec]{INFOTEC Centro de Investigación e Innovación en Tecnologías de la Información y Comunicación. Cto. Tecnopolo Sur No. 112, 20326 Aguascalientes, Ags. México.}
\address[centrogeo]{CentroGEO Centro de Investigación en Ciencias de Información Geoespacial. Circuito Tecnopolo Norte, No.107 Col. Tecnopolo Pocitos II, 20313 Aguascalientes, Ags. México.}

\begin{abstract}
Finding similar objects in a collection of high dimensional data through a distance function is a pervasive task in computer science and many related fields like machine learning and information management and retrieval. The generic pipeline preprocesses the data collection to produce an index that allows the efficient retrieval of nearest neighbors. Nonetheless, high-dimensional data will suffer from the so-called {\em curse of dimension} (CoD). In this scenario, the CoD is observed as the impossibility of creating efficient algorithms to retrieve the nearest neighbors for given queries in a collection. The strategy to overcome slow computation is trading quality by search speed, i.e., algorithms are allowed to retrieve elements that must not be in the exact results and ignore some that must be in the result set. Applications that require high-quality results but cannot afford the required computational costs will require algorithm experts for tuning indexes and searching algorithms to obtain the desired speed and quality.

This manuscript introduces an autotuned algorithm for constructing and searching nearest neighbors based on neighbor graphs and optimization metaheuristics to produce Pareto-optimal searches for quality and search speed automatically; the same strategy is also used to produce indexes that achieve a minimum quality. Our approach is described and benchmarked with other state-of-the-art similarity search methods, showing convenience and competitiveness.
\end{abstract}


\end{frontmatter}


\section{Introduction}
\label{sec/intro}

The \emph{proximity search} problem can be described in terms of \emph{metric spaces} which have a solid theoretic base. This abstraction allows the problem to be applied to a wide variety of fields like Statistics, data mining, pattern recognition, multimedia information retrieval, machine learning, bioinformatics, and others \cite{Ssisap2010}. The problem definition is succinct, given a metric space $(U, d)$ and a subset $S=\{s_1,\ldots,s_n\}$ of $U$, called the database, the problem of similarity search is how to preprocess $S$ to find objects close to a query $q$ for every possible $q \in U$ under metric $d$. The objects close to a query $q$ can be defined by $B_r(q)=\{x\in S \mid d(x, q)\leq r\}$ if $r$ is given or $\knn(q)$ which contains the $k$ nearest neighbors of $q$ in $S$, if the number $k$ is given. In this work, we focus on the latter.

Finding the $\knn$ can be solved by brute force by simply evaluating the entire dataset looking for the nearest neighbors. When the database is so extensive, this naive solution takes too much time to be useful. In many problems, it is reasonable to sacrifice the exactness of the search in order to boost its speed. In these scenarios, the \emph{recall} is often used to measure the quality of the search.

The data structure produced by preprocessing the database is called the \emph{index}. It is possible to create an index that enables searches with high recall scores for a \emph{simple} database. The complexity of a database can be measured by its \emph{intrinsic dimension} which can be observed as a high-concentration around the mean of all paired distances \cite{CNBYMacmcs2001}. It is well known that the higher the intrinsic dimension, the harder it is to build an index that will produce searches with high recalls. This phenomenon is known as the \emph{curse of dimensionality}.

\subsection{Related work}

{\em Locality Sensitive Hashing} (LSH) is among the most well-known techniques for approximate similarity search. It was introduced in~\cite{GIMvldb1999}, with many follow up work as in \cite{AIacmcomm2008,heo2015spherical,andoni2015practical}. An LSH method defines a set of ad-hoc hashing functions that partition the database with the property that close elements of $S$ are mapped (with high probability) to the same bucket and far away elements are mapped (again, with high probability) to different buckets. In this way, the search can be done only on the most promising buckets avoiding unnecessary comparisons.

A different approach is made by methods that use a small collection of elements of the metric space (often called references) to project each element of the database to an alternative space. In particular, a family of methods based on the rank of references as perceived by objects has been proved to be of use in high dimensional datasets~\cite{CFNieeepami2008}. The approach can also be applied to large datasets if each object focuses on the $k$ nearest references since an inverted index or a trie can accelerate the searches. Some examples of this method are CNAPP \cite{TCNis2012}, PP-Index \cite{Eppindex2014}, MIF \cite{AGSmetric2014}, and the quantized permutations~\cite{MOHAMED2015}. The structural similarity was systematically explored in \cite{KNR2015}, adding several indexes to the list. Also, a detailed exploration on the choice of the references is on \cite{AMATO2015}.

A popular approach for finding the elements of a database that produce the higher inner product to the query is the \emph{quantization} method \cite{jegou2011}. These methods use two main strategies. On the first hand, they partition the space to reduce the computational cost of solving queries. The partition is usually in buckets, such that the algorithm that solves a query can discard most buckets, and only a few promising buckets need to be evaluated. The second technique quantizes the data to produce faster (and not exact) product comparisons. The quantized data is more compact, improving the computation times due to the memory hierarchy.

On \cite{avq_2020} the authors presented SCANN, a new method based on optimizing a loss function for quantization methods that takes into account the dot product of the elements. Points with higher dot products with the query set are preferred when minimizing the loss function. In the experiments, the authors showed that the proposed loss achieves better recall and inner product value estimation when compared with losses that focus only on minimizing the reconstruction error, as other product quantization methods do.

\subsection{Neighbor graphs}
\label{sec/navigating}
When the database is transformed into a graph, where edges reflect metric space's distances, the search can be done by navigating the nodes using the edges such that the distance to the query gets reduced on each step. The most popular algorithm for navigation is the greedy search, where the next path is the immediately most promising. On \cite{10.14778/3303753.3303754}, Fu et al. used this search algorithm on an approximation of the graph presented in Spatial Approximation Tree (SAT) \cite{nav02}. The approximation is made with a starting nearest neighbor graph produced by methods like NNdescent (see below), and then it improves it by taking the new neighbors using the graph from SAT. Using this graph, they created the NSG index, where they found empirically that the search time is close to $O(\log n)$.

The DISK-ANN \cite{NEURIPS2019_09853c7f} is an index designed to be stored in an SSD, which means that the number of accesses is minimal (like a dozen). The authors used a graph-based index for doing the searches. They created the graph to find the nearest neighbor using the greedy search algorithm. They used a similar idea of the Spatial Approximation Tree (SAT) \cite{nav02}; the algorithm was modified to discard space's regions in the form of hyperbolas instead of hyperplanes. The DISK-ANN is built on SSD's clusters, where each element in the database is assigned to the $\ell$ closest clusters. The distance comparisons are approximated using quantization techniques \cite{jegou2011}.

In \cite{10.1007/978-3-319-46759-7_2}, Iwasaki presented the case where an NNG (Nearest Neighbor Graph) with more edges is going to be bigger, and the searches are going to be slower. That is why they developed a method to prune some of the edges. First, they showed how to get a bidirectional graph from a directed one. They add the edges in the other direction, and if the number of edges surpasses a limit, they remove the longest edges. That is because, the author said, long edges do not contribute to finding the actual nearest neighbors. This strategy is used over an approximate NNG to create NGT-PANNG where all the searches start at the same initial node. 
Latter, in \cite{DBLP:journals/corr/abs-1810-07355} Iwasaki et al. paid attention to both the indegree and the outdegree of the nodes in the graph since they found out that, given an NNG, if the edges on the nodes with an outdegree of 40 are reversed, the query time is reduced. The authors presented a method to adjust the indegree and the outdegree for each node. Also, they presented a path adjustment where they removed shortcuts if there is an alternative path of two nodes and these two nodes are shorter than the edge in question. In addition, they showed how to dynamically adjust the indegree and outdegree during search time to let the user decide the trade-off between precision and speed.

In \cite{10.1145/1963405.1963487}, Dong et al. presented NNdescent, a new heuristic for constructing an approximated NNG based on the observation that a neighbor of a neighbor is also likely to be a neighbor. This principle applies to metric spaces that are \emph{growth restricted}, that is, there should exist a constant $c$ such that $|B_{2r}(x)|\leq c|B_r(x)|$ for all $x\in S$. The heuristic to find the nearest neighbors is as follows. First, a coarse approximation of the nearest neighbors for every $x$, $B(x)$ is created from $k$ random elements. The next steps are going to produce better approximations to $B(x)$ by exploring the neighbors of every $y\in B(x)$ (including the reverse neighbors). The growth-restricted property, together with independence assumptions, guarantee that the radios of every $B(x)$ will be reduced by half on each iteration. This process ends when convergence. In the end, the NNG is going to be a list of all the approximations $B(x)$ of the real neighbors for every $x\in S$.

An alternative approach is made by the Navigable Small World (NSW) index~\cite{MPLK14}. The core idea is to create a graph of $k$ approximate nearest neighbors and use it to search. The search procedure greedily navigates the graph following those nodes that minimize the radius of the result set populated as the algorithm navigates the graph. The graph is constructed with the search algorithm since inserting the $i$th element implies searching nearest neighbors in the previously indexed $i-1$ elements. The rest of this section reviews the algorithm and some variants used here.

The NSW construction requires a distance function $\dist$ and a database $S = u_1, u_2, \cdots, u_n$, $S \subset U$. The core idea is to create the search graph $(S, \N)$, where vertices are the objects of the dataset and the transition function $\N$ represents the set of edges. That is, $\N_u$ is the adjacency list of $u$. The construction is incremental using the following rules:
\begin{itemize}
    \item The graph is empty, $(\{\}, \{\})$ and adds $u_1$: the graph is updated as $(\{u_1\}, \{u_1 \rightarrow \emptyset \})$
    \item The graph $(S^{(i-1)}, \N^{(i-1)})$ is not empty and adds $u_i$ (the $i$th element): the graph is then updated to $(S^{(i)}, \N^{(i)})$ where $S^{(i)} = S^{(i-1)} \cup \{u_i\}$. To update the neighborhood function, we firstly compute the approximate nearest neighbors $\knn(u_i, S^{(i-1)})$ using the search algorithm on $(S^{(i-1)}, \N^{(i-1)})$.
    The new $\N^{(i)}$ is defined as 
    $\{u_i \rightarrow k\textsf{nn}(u_i, S^{(i-1)})\} \cup \N^{(i-1)}$
    but also adding $u_i$ as neighbor to all $\N^{(i-1)}_v$ for $v \in \knn(u_i, S^{(i-1)})$, i.e., adds $u_i$ as reverse neighbor into the adjacent list of its nearest neighbors. The number of neighbors $k$ is a construction parameter.
    \item Neighborhood reduction: In~\cite{tellez2021scalable} the set of nearest neighbors $k\textsf{nn}(u_i, S^{(i-1)})$ is also reduced using a Spatial Approximation Tree (SAT) step \cite{nav02}, i.e., let $X = k\textsf{nn}(u_i, S^{(i-1)})$ so $X=(x_1,\ldots,x_{|X|})$ are the ordered neighbors of $u_i$, the SAT reduced set $X^\text{SAT}$ centered in $u_i$ is constructed incrementally as $X^\text{SAT} = \{x_1\}$ and then, for $j=2$ to $|X|$, $x_j \in X^\text{SAT}$ if there is no other object in the current $X^\text{SAT}$ that is nearest to the center object $u_i$, i.e., $\dist(u_i, x_j) < \min \{ \dist(v, x_j) \mid v \in X^\text{SAT} \}$. This produces smaller neighborhoods that are shaped by the data's distribution (the $X^\text{SAT}$ will stop growing when the centered object is \textit{covered} an there is no other object that can be inserted into it) instead of a given parameter.
    \item Neighborhood size $k$: instead of specifying a fixed number, as in the NSW and HNSW~\cite{malkov2018efficient}, we use the variable strategy presented in~\cite{tellez2021scalable}, where a logarithmic neighborhood size is used, i.e., $k=\lceil \log_b(n) \rceil$. In this approach, the logarithmic base $b > 1$ is used, and it is intended to control the memory of the index. In particular, we search for logarithmic neighborhoods ($1 < b \leq 2$) and then apply SAT reductions to obtain competitive indexes regarding memory, search time, and quality.
\end{itemize}

Please note that the construction algorithm is tightly coupled with the search algorithm. The nearest neighbor query $k\textsf{nn}(q, S)$ is solved by a search algorithm. In particular, the NSW is an iterative procedure starting at a random point $u$ and greedily navigating the graph using its neighborhood to reduce the radius of the $k$th approximate nearest neighbor until it is impossible to improve. The procedure is repeated several times to improve the result set, i.e., the number of repetitions is a hyper-parameter. The HSNW~\cite{malkov2018efficient} improves the algorithm introducing a hierarchical structure to navigate the graph more efficiently, in contrast, the construction and searching algorithms becomes more complex. Our previous work~\cite{tellez2021scalable} uses a Beam Search (BS) variant for searching. The BS is a one-pass population-based algorithm where its main hyper-parameter is the size of the population $\bsize$; the population is called the \textsf{beam}. Another hyper-parameter used in the same work is the initial population size $\ssize$, typically larger than the size of the $\bsize$. The algorithm navigates the graph tracking best known $\bsize$ items; these elements are refined while revising each note and are also used to guide the navigation. The result set is refined during the whole process; Section~\S\ref{sec/bs} details the algorithm along with our new proposals.

\subsection{Overview}
This section describes the similarity search problem and surveys the related literature. The rest of the manuscript is organized as follows. Section~\ref{sec/methodology} describes our methodology and benchmarks.
Section~\ref{sec/bs} revisits the similarity search index based on the Beam Search metaheuristic; it covers both our searching and construction contributions based on automatic tuning of hyperparameters. The parameter space used for our automatic tuning is described in detail in Section~\ref{sec/hyper-parameter-optimization}; this section also covers our three error functions used in the optimization. Section~\ref{sec/experimental-results} characterizes and compares experimentally our contribution and state-of-the-art alternatives. Finally, our manuscript is summarized, and a list of future research lines is also discussed in Section~\ref{sec/conclusions}.

\subsection*{Our contribution}
This manuscript introduces several improvements to the BS algorithm: a new search hyper-parameter $\Delta$ that can be used effectively to trade speed and quality of the searches, automatically determine search hyper-parameters using meta-heuristics, in particular, we use the same Beam Search algorithm used for searching and a genetic-based algorithm to perform this hyper-parameter optimization task. Using this framework, we propose three different optimization strategies: \textsf{Pareto-recall}, \textsf{Pareto-radius}, and \textsf{Min-recall}. The first two aim to find the best trade between quality and search speed, and the second one is the best trade ensuring a given minimum quality. Finally, the construction algorithm is also modified to take advantage of these strategies and keep the number of hyper-parameters low without compromising speed or quality.

\section{Methodology}
\label{sec/methodology}

Our auto-tuned algorithms are motivated by experimental observations, and therefore, we will describe and characterize our algorithms using experimental evidence. We also compare our approach with state-of-the-art methods using a similar methodology.
Table~\ref{tab/benchmarks} lists the set of benchmarks and their characteristics such that they contextualize our experimental results.

\begin{table}[!ht]
\centering
\caption{Benchmarking datasets. We list their train (dataset) and test (queries) sizes, its measure distance, and their dimension.}
\label{tab/benchmarks}

\resizebox{0.8\textwidth}{!}{
\begin{tabular}{rr rcr}
\toprule
\bf name &\bf train &\bf test &\bf distance &\bf  dimension \\
         &\bf $n$       &\bf $m$        &\bf function &               \\
\midrule
  BigANN-100M  & 100,000,000  &  10,000  & $L_2$  & 128 \\
    BigANN-1M  &   1,000,000  &  10,000  & $L_2$  & 128 \\
DeepImage-10M  &   9,990,000  &  10,000  & $\cos$ &  96 \\
      GIST-1M  &   1,000,000  &   1000   & $L_2$  & 960 \\
   Glove-1.1M  &   1,183,514  &  10,000  & $\cos$ & 100 \\
   Glove-400K  &    390,000   &  10,000  & $\cos$ & 100 \\
  Lastfm-300K  &    292,385   &  50,000  & $\cos$ &  65 \\
   Twitter-2M  &   2,262,034  &  10,000  & $\cos$ & 300 \\
     WIT-300K  &    308,374   &  10,000  & $\cos$ & 512 \\
\bottomrule
\end{tabular}
}
\end{table}

Our cost-measuring methodology consists of various benchmarks that compare search and quality of the result set, measured as $\recall$, the time costs in seconds for the construction and the search time,\footnote{We also uses the number of distances evaluations when this is specified.} and finally, we also report memory usage as megabytes. Let us properly define our quality measure, the recall:
\[ \recall = \frac{\# \text{ of real } k \text{ nearest neighbors}}{k}, \]
along with the manuscript, we present recall scores as the macro-recall of all queries in a benchmark, i.e., the mean of all recall values per query. We fixed $k=32$ for all our experiments, and our gold standard was computed using an exhaustive evaluation.

We index a {\em dataset} (train partition of size $n$) and measure the performance in a set of queries (test partition of size $m$). Please ignore the abuse of train and test terminology, but it helps clarify our methodology. We never touch the test partition during the indexing phase and also never touch the test while optimizing hyper-parameters. 

We selected our benchmarks to be public and its measures used in both real applications and research works.
We use the last seven datasets of the table to characterize our methods; these datasets range from close to 300K (Lastfm-300K) objects to close to 10M objects (DeepImage-10M). The query sets range from one thousand objects (GIST-1M) to 50K queries (Lastfm-300K). Our benchmarks are vector datasets using different dimensions, ranging from $65$ (Lastfm-300K) to 960 (GIST-1M). While our methods are general enough to work with other representations and distance functions, we use benchmarks that work under $L_2$ and $\cos$ distances. Among the seven characterization datasets, two of them (Twitter-2M and WIT-300K) were introduced by us.\footnote{The Twitter-2M and WIT-300K datasets and a list of demonstrations are available at \url{https://github.com/sadit/SimilaritySearchDemos}. }

Our methods never touch \textsf{test} partitions during the indexing procedure, but we also show that the empirical observations of algorithms and meta-heuristics still apply on two samples of 1 and 100 million datasets of the so-called BigANN dataset, see Table~\ref{tab/benchmarks}. These datasets are not observed during the characterization process. They are used as an experimental verification that our methods apply for other datasets with at least an order of magnitude larger than the rest of our benchmarks.

We ran our experiments in an Intel(R) Xeon(R) Silver 4216 CPU @ 2.10GHz workstation with 256GiB RAM using GNU/Linux CentOS 8. Our system has 32 cores with hyperthreading activated (64 threads). Our algorithms were implemented in the Julia language (v1.6.3); we used 64 threads for construction and a single thread for searching. Finally, we remark that our implementation is a registered Julia package called SimilaritySearch.jl\footnote{\url{https://github.com/sadit/SimilaritySearch.jl}} licensed under the MIT open-source license.

\section{Revisiting beam search algorithm over neighbor graphs}
\label{sec/bs}

\begin{algorithm}[!ht]
\small
\begin{algorithmic}[1]
\Require The distance function $\textsf{dist}$, the search graph $G=(S,\N(S))$, the query $q$, the number of neighbors $k$, and the hyper-parameters: the beam size $\bsize$, the expansion factor $\Delta$, and the number of maximum visits $\maxvisits$.
\Require Define objective function as $\fitness(u) = \dist(q, u)$ for the similarity search task.
\Ensure The result set $R$ containing $k$ approximate nearest neighbors of $q$.
\Function{BeamSearch}{$\fitness, S, \N, k, \bsize, \Delta, \maxvisits$}
\State initialize $B$ and $R$ as priority min-queues of size $\bsize$ and $k$, respectively.
\State $\states \leftarrow \emptyset$ \Comment{a set with \textsf{visited} vertices}

\State populate $R$ and $\states$ \Comment{starting points in $R$ also populate $\states$}
\State add $(\mathsf{min}(R), \mathsf{argmin}(R))$ into $B$ \Comment{seeds $B$ with the current best}
\While{$|B| > 0$} \Comment{explores iteratively until $B$ is empty}
    \State $p \leftarrow \textsf{popmin}(B)$ \Comment{removes best approximation in $B$}
    \For{$c \in \N(p)$} \Comment{visit each child in the neighborhood of $p$}
        \If{$c \not\in \states$} \Comment{ignores already visited elements}
            \State add $c$ into $\states$ \Comment{marks $c$ as visited}
            \State add $(\fitness(c), c)$ into $R$ \Comment{$R$ accepts adds if $\fitness(c) \leq \mathsf{max}(R)$ or $|R| < k$}
            \If {$|\states| = \maxvisits$}
                \State \Return $R$ \Comment{early stop search procedure}
            \ElsIf{$\fitness(c) \leq \Delta \cdot \mathsf{max}(R)$}
                \State add $(\fitness(c), c)$ into $B$ \Comment{$B$ accepts adds if $\fitness(c) \leq \mathsf{max}(B)$ or $|B| < k$}
            \EndIf
        \EndIf
    \EndFor
\EndWhile

\State \Return $R$
\EndFunction
\end{algorithmic}
\caption{Beam search algorithm.}
\label{alg/bs}
\end{algorithm}

We use the Beam Search (BS) algorithm as the basis for our contribution; see our previous work~\cite{tellez2021scalable}. Alg.~\ref{alg/bs} details our new approach, which is a simplification of the previous algorithm and also provides support for the new parameter $\Delta$. The BS algorithm starts populating a priority queue $R$, which will be the final approximation of the $k$ nearest neighbors when the algorithm finalizes. Line 5 takes the best element in $R$ to initialize the beam $B$. 
Then, the main loop (lines 6-19) corresponds to the main search procedure that repeats while $B$ has elements (or it is early stopped in line 13). The body of the main loop contains another loop (lines 8-18) that explores the neighborhood of the best-known approximation. Vertices are only visited once (ensured by tracking visited vertices with $\states$). Line 11 describes the main push to the result set $R$ (the properties of this min-queue structure are detailed in the following paragraphs). Condition of line 14 checks if objects should be considered by the beam, i.e., using the $\Delta$ parameter as expansion factor of the maximum acceptable fitness value in $R$. The $\Delta$ factor indirectly controls the permanency in the main loop (lines 6-19).

Alg.~\ref{alg/bs} presents several changes with respect to that found in~\cite{tellez2021scalable}. Firstly, it was simplified and now we use a $\fitness$ function instead of $\dist$ evaluations. Secondly, it uses the $\Delta$ factor that controls the population of the beam $B$; $\Delta=1$ is equivalent to the original algorithm. Thirdly, we introduced the $\maxvisits$ parameter that provides a mechanism for early stopping the search, a property needed for fast exploration of hyper-parameter spaces; $\maxvisits=|S|$ reproduces to the original behavior.

Our algorithm has several requirements, more detailed:
\begin{itemize}
    \item The $\fitness : U \rightarrow \mathbb{R}^+$ function to be minimized.
    \item $\fitness(u) = \dist(q, u)$ for finding nearest neighbors of $q$, both $q$ and $u$ are valid object.
    \item The $\N_u$ set containing the neighborhood of $u$ (linked vertices in the graph).
    \item An initialization process for the result set (line 4), our similarity search procedure uses a static sample of objects of $S$; the sample is updated on exponential steps while inserting elements with the construction process.
    \item Priority queues of maximum length $k$ (\textsf{p-queue} for short) store pairs $\mathbb{R}^+ \times U$.
\end{itemize}

One main piece of our algorithmic design is the p-queue, a min-max priority queue that limits its maximum size. Proper implementation and analysis are beyond the scope of this manuscript. However, the p-queue $R$ has the following operations and properties:
\begin{itemize}
    \item The $\mathsf{min}(R)$ and $\mathsf{max}(R)$ functions defined as the minimum and maximum $\fitness$ values in $R$.
    \item The $\mathsf{argmin}(R)$ function is defined as the object associated with the minimum $\fitness$ in $R$.
    \item The $\mathsf{popmin}(R)$ and $\mathsf{popmax}(R)$ functions remove the pair associated respectively with the minimum and maximum $\fitness$ values in $R$, both return the removed object.
    \item It accepts adding pairs $(\fitness(u), u)$ if $\fitness(u) \leq \mathsf{max}(R)$ or $|R| < k$, for $k$ being the maximum size of $R$.
    \item If a pair $(\fitness(u), u)$ is accepted for insertion and $|R| = k$ the queue's size must remain $|R| = k$ after the operation, e.g., applying $\mathsf{popmax}(R)$ before the actual insertion.
\end{itemize}

Our implementation consist of two arrays sorted by priority, i.e., solves $\mathsf{min,max,argmin}$ in $O(1)$ time. We also use a floating pointer for tracking minimum value ($\mathsf{popmin}$ and $\mathsf{popmax}$ in $O(1)$ time). The insertion algorithm is simple, a key-value fast insertion-sort-based algorithm for adding accepted pairs. Nonetheless, it takes advantage of the expected $\fitness$ values distributions, i.e., new $\fitness$ values will not achieve insertions since new probes will likely not improve the result set. The p-queue allows describing our search algorithm in simple terms.

\subsection{Construction algorithm}
\label{sec/construction}

This manuscript modifies the searching algorithm to support new hyper-parameters and modifies the construction algorithm to reduce user intervention in the tuning process. Here we revisited the construction algorithm, using as starting point the algorithm and modifications listed in \S\ref{sec/navigating}.
For this matter, we report the following improvements:
\begin{itemize}
\item An automatic hyper-parameter tuning based on the beam search algorithm, we define a search space and three ad-hoc error functions.
\item A strategy to adjust hyper-parameters that supports incremental construction.
\item A parallelized construction algorithm that takes advantage of modern multi-core architectures.
\end{itemize}
These modifications are compared with state-of-the-art indexes in the experimental section.

\subsubsection{Hyperparameter optimization}
Our search algorithm introduces the $\Delta$ parameter, which allows controlling the graph's exploration. We also kept the $\bsize$, limiting the population's size in the queue that the search process will visit. These hyperparameters must be adjusted to take advantage of the algorithm for a given task. This problem is common to any search index, and it is a barrier that limits the broad usage of efficient algorithms. We propose an online model selection procedure that optimizes the hyper-parameters by reducing a given error function. The optimization is performed in the same terms as our search graph; we use the beam search algorithm on a specially defined graph over a space of configurations $(\bsize, \Delta)$ pairs. The error function to be minimized is the objective of any search algorithm: speed and recall. The model selection procedure is detailed in \S\ref{sec/hyper-parameter-optimization}, where the configuration space is explained, and the error functions are also introduced.

\subsubsection{Maintaining hyper-parameters}
The incremental construction of the search graph allows the index to solve queries at any time. However, the hyper-parameters selection is made before the index's construction, and its effectiveness is tested after the build. Thus, the requirements at any construction stage could be quite different, and setting hyper-parameters that will work on large datasets may result in higher construction times. In this sense, we run a hyper-parameter optimization $\lceil \log_b{n} \rceil$ for a dataset of $n$ objects, in particular, we adjust parameters when $\lceil \log_b i \rceil + 1 = \lceil \log_b{(i+1)} \rceil$. 
Moreover, we use the same policy for computing the initial approximation of the result set for the beam search, see line 4 of Alg.~\ref{alg/bs}. Instead of our previous approach, i.e., initializing the result set with a random sample, here we select a set of objects where for each pair of elements $u, v$ on it we ensure that $\N_u \cap \N_v = \emptyset$. We use a logarithmic sample size, $\lceil \log_b i \rceil$, to simplify the selection of parameters. 
The cost of these logarithmic adjusts, and the computation of the initial approximations is amortized effectively with the entire construction cost. The number of tuning processes is also enough to impact the results. We verified these claims experimentally in \S\ref{sec/experimental-results}.

\subsubsection{Taking advantage of multi-threading architectures}
We use a parallel algorithm to reduce construction times. We preserved the incremental construction since it can be of interest on many applications domains to have a fully functional index at any construction stage.
We tackled the parallelism using block processing; that is, we select at most \textsf{block\_size} items that are inserted at a time. The algorithm is quite similar to the original one, see \S\ref{sec/intro}, yet we add some requirements:
\begin{itemize}
    \item If the graph is empty or contains less than \textsf{block\_size} elements, then a sequential insertion is made.
    \item Given the graph $(S^{(i)}, \N^{(i)})$ and a block of $\mathsf{block\_size}$ elements. For each $u$ in the block, we search $\knn(S^{(i)}, u)$, searches are performed in parallel. The neighborhood $\N_u$ is created; please recall that the SAT algorithm is also applied to reduce the set of approximate nearest neighbors. Note that the computations made for $u$ can be performed using the same thread. Once all searches and direct neighborhoods are computed, the reverse connections are then performed in another block of threads to avoid possible data races when searching. In this sense, we also lock $\N_v$, $v \in S^{(i)}$, since each $v$ can be in more than one set of nearest neighbors in the block. The new graph is then defined, including all items in the block.
\end{itemize}

We tested the parallel block size and found that it is enough to select it several times larger than the number of available threads. Small block sizes are less effective than larger ones until it converges when block sizes are set to several thousand elements. Nevertheless, if the block size is too large, then the effect of ignoring elements in the same block may be noticed. We fix the block size to $1024$ since our experiments indicate that this value is good enough to avoid a noticeable quality's impact and also being after the faster growth rate in the $\mathsf{block\_size}$ vs. {\em construction time} curve. The experiments supporting our decisions are not presented in this manuscript to keep it concise and straightforward. Nonetheless, all experimental results presented in the rest of this document argue its viability. We calibrated \textsf{block\_size} using the Glove-400K dataset in our 32 core system. 

\section{Hyperparameter optimization with the Beam Search}
\label{sec/exploring-parameter-space}
It is well-known that there is a trade-off in metric search algorithms and their associated indexes; it is hard to obtain high-quality, high-speed searches and also keep low memory requirements. Fine-tuning search algorithms for real datasets is a computationally expensive task, not to mention the time and the stepped learning curve required by the people doing the job.

\begin{table}[!ht]
\caption{BS memory and time construction costs of our benchmarks for different neighborhood sizes ($b=2.0, 1.5, 1.2, 1.1$), beam size of 32 ($bsize$), without search pruning or expansion, i.e., $\Delta=1$. The presented time-costs were obtained using 64 threads. The memory measure also includes the necessary RAM to store each dataset, i.e., $n$ 32-bit floating-point vectors of dimension $m$.}
\label{tab/construction}
\centering

\resizebox{\textwidth}{!}{
\begin{tabular}{r rrrr rrrr}
\toprule
     & \multicolumn{4}{c}{\bf memory (MiB)} & \multicolumn{4}{c}{\bf construction time (sec.)} \\ \cmidrule(lr){2-5} \cmidrule(lr){6-9}
\bf dataset    & $b=2.0$ & $b=1.5$ & $b=1.2$ & $b=1.1$ & $b=2.0$ & $b=1.5$ & $b=1.2$ & $b=1.1$ \\ 
\midrule
DeepImage-10M  &  4,517  & 4,727   & 5,064    & 5,398  &  572 & 390 & 269 & 195 \\
      GIST-1M  &  3,715  & 3,725   & 3,738    & 3,751  &  214 & 152 &  95 & 62  \\
   Glove-1.1M  &    545  &  575    & 632      &   698  &   69 &  39 &  22 & 17  \\
   Glove-400K  &    176  &  182    & 182      &   203  &    9 &   6 &   4 & 13  \\
  Lastfm-300K  &     88  &   90    &  92      &    95  &    3 &   2 &   1 & 11  \\
   Twitter-2M  &  2,744  &  2,779  & 2,840    & 2,909  &  278 & 173 & 100 & 70  \\
     WIT-300K  &    625  &  629    &  634     &   639  &   18 &  12 &   9 &  8  \\
\bottomrule
\end{tabular}
}
\end{table}

\begin{figure}
    \centering
    \includegraphics[width=\textwidth,trim={0cm 23cm 0 0},clip]{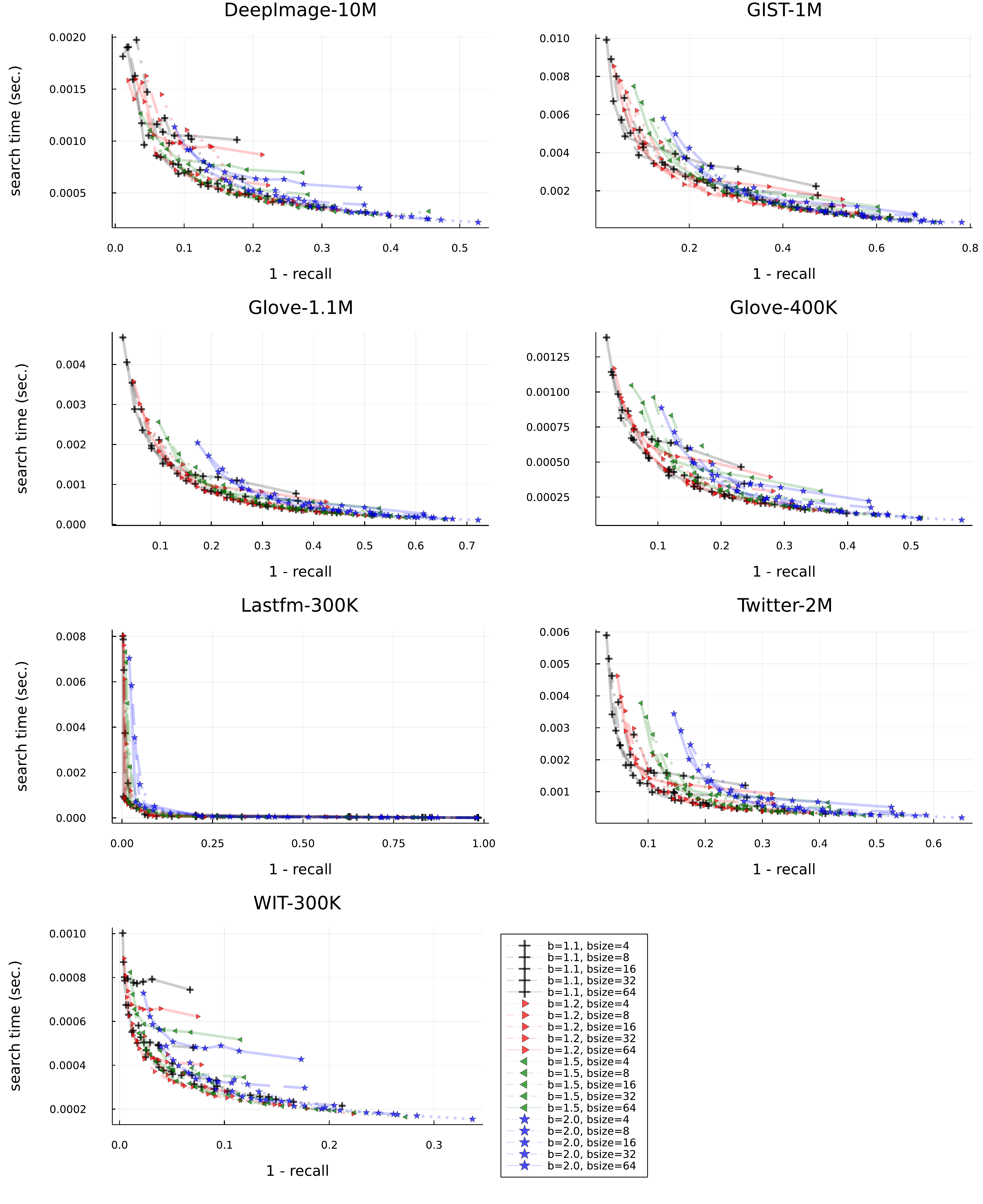}
    \caption{Performance regarding both search and $\recall$ across benchmarks for different parameters of $b=1.1, 1.2, 1.5, 2.0$ and $\bsize=4, 8, 16, 32, 64$; each curve corresponds to different values of $\Delta=0.9, 0.95, 0.97, 0.99, 1.0, 1.05, 1.1, 1.15, 1.2$. Points and curves closer to the origin are better since means smaller search times and smaller errors (high $\recall$).}
    \label{fig/search-delta}
\end{figure}

Figure~\ref{fig/search-delta} illustrates the performance of our modified BS. The figure shows how varying $b$, $\Delta$ and $\bsize$ it is possible to explore the hyper-parameter space to create indexes that trade memory, search-cost, and $\recall$ ($1 - \recall$ to be precise, which can be seen as an error to be minimized). 
In particular, the indexes shown in this figure were created using logarithmic neighborhoods with SAT reduction (logarithmic bases $b=1.1, 1.2, 1.5,$ and $2.0$) using $\bsize=32$ and $\Delta=1$. The construction times and memory usage used per each index are listed in Table~\ref{tab/construction}. Both $\bsize$ and $\Delta$ parameters were changed in a grid for search time, and each curve in the figure represents the performance for the same $b$ (same $b$ kept same markers and colors) and $\bsize$ (same $\bsize$ kept same line types), with varying $\Delta$ values (points in the same line, we applied small changes in its values from 0.9 to 1.2). We can observe how the $\Delta$ parameter changes the search performance significantly, using the same index, even surpassing the positive impact of $\bsize$. For instance, please observe continuous lines, i.e., $\bsize=64$, we can appreciate clearly in DeepImage-10M, Twitter-2M, and WIT-300K, how these curves increase the search time significantly. Smaller $\bsize$ values achieve similar result qualities without negatively impacting the search speed. On the other hand, increasing $\Delta$ also increases the search time, but it can produce much larger improvements regarding $\recall$. Regarding $b$, it is also possible to observe that smaller values (i.e., larger indexes) can achieve faster searches and higher quality results. For instance, this is observed by black and red curves being closer to the vertical axis, and contrarily, green and blue are distant to the vertical axis. 

It is worth mentioning the hyperbolic shape in Figure~\ref{fig/search-delta} curves; please recall that points closer to the origin are search configurations that produce faster searches with high-quality results. Points closer to the vertical axis are also of particular interest since they may produce acceptable configurations for those applications needing high-quality results, even when they may not be as fast as other setups.

Space's parameter exploration is necessary to achieve affordable computational costs with acceptable quality. Nonetheless, the exploration requires the evaluation of hundreds of configurations and proficiency in the particular algorithms that impose a barrier for using these methods.

\subsection{Genetic-based exploration of the configuration space}
\label{sec/hyper-parameter-optimization}
Here we introduce a method for automatically tuning search hyper-parameters of our BS algorithm for similarity search. Our goal is to reduce the complexity of using our similarity search algorithm to almost a unique decision: selecting the memory of the index (the parameter $b$ used for constructing the index). The procedure explores BS's $\bsize$ and $\Delta$ parameter space to achieve competitive speeds and qualities. Please note that the exploration itself has several parameters, as discussed in this section. However, its final description is quite generic and does not need to be adjusted for typical workloads, as experimentally shown in \S\ref{sec/experimental-results}.

A noticeable property of our approach is the usage of the beam search itself to tune hyper-parameters, see Alg.~\ref{alg/bs}. As explained in \S\ref{sec/bs}, beam search is a population-based algorithm that iteratively tries to reduce a cost function visiting nodes in a graph of neighbors. The main idea is to explore neighborhoods while keeping a small set (\textit{beam}) of promissory vertices to be explored. 
In contrast with the similarity search task, the hyper-parameter search graph $H=(C, \N)$ is not explicitly stored; it is defined as follows: $C = \mathbb{N} \times \mathbb{R}^+$ which is the domain of $\bsize$ and $\Delta$; in this sense $\mathcal{N}: C \rightarrow C^\ell$ where $C^\ell$ is a subset of $C$ of size $\ell$. Therefore, the task becomes bounding $C$, the precise $\N$ definition, and a properly $\fitness$ function to drive the navigation.

\subsection{Bounding the parameter space $C$}
As commented, each $c \in C$ is a tuple with instances of hyper-parameters $(\bsize, \Delta)$ also called configurations. Please recall that $\bsize$ is the beam size that limits the maximum number of elements that are kept at any stage of the search, see \S\ref{sec/bs}; then, it is an unbounded positive integer, but its useful range is relatively small. We limit $\bsize$ to be between $2$ and $512$. Figure~\ref{fig/search-delta} shows the performance values between 4 and 64, and they work for most of our benchmarks. On the other hand, $\Delta$ is an expansion parameter that allows considering more or fewer elements to be part of the beam at each iteration. It controls if the algorithm stops even without converging or remains exploring even if it locally converges, see Alg.~\ref{alg/bs}.
The original BS Search Graph fixes $\Delta=1.0$. While useful values are determined by the underlying distance distribution, we limit $0.6 \leq \Delta \leq 2$ since two balls of the same radius $r$ will not intersect each other if $d(u, q) > 2r$. This heuristic is not a rule since the graph is not a partition, and we cannot precompute a precise limiting radius. Figure~\ref{fig/search-delta} show that restricting $\Delta$ to a less wide range, $0.9 \leq \Delta \leq 1.2$, it is possible to trade effectively search speed and quality, i.e., reach close to perfect $\recall$ in all benchmarks and can also obtain high-speed searches.

\subsection{Neighborhood $\N$}
The neighborhood function connects each configuration with other configurations that ideally are similar in some sense. While the precise similarity notion is not relevant for our approach, we will define how to construct neighborhoods. Inspired by genetic algorithms, we define three different functions on $C$ and configurations $c \in C$:
\begin{description}
\item[$\rand(C)$:] samples the configuration space $C$. Our implementation selects two valid random values for $\bsize$ and $\Delta$.
\item[$\mutate((\bsize, \Delta), C)$:] creates a new configuration modifying the previous one, following limits in $C$. In particular, our implementation creates a configuration in the following way $(\scale_p(\bsize, \alpha), \scale_p(\Delta, \beta)$ where $\scale_p(a, b)$ is a function that with probability $p$ computes $ab$ and $a / b$ with probability $1-p$. The lower and upper valid bounds of each hyper-parameter must be ensured.
\item[$\crossover((\bsize_1, \Delta_1), (\bsize_2, \Delta_2))$:] combines two configurations into a new one; our implementation computes $(\lceil(\bsize_1 + \bsize_2) / 2\rceil, (\Delta_1 + \Delta_2) / 2)$.
\end{description}

We are now able to define the initialization procedure for BS (Alg.~\ref{alg/bs}, line 4) as a random sample of $C$, using $\rand(C)$. Moreover, $\N(c)$ is defined as $\{ \mutate(c) \}^{\gamma}$; we also include $\delta$ configurations computed with $\crossover$ over two configurations randomly selected from the beam. The latter adds some diversity to the neighborhood, trying to avoid local optima points, i.e., similar to long links mentioned in \cite{MPLK14}.

\subsubsection{Our implementation setup}
Our actual implementation limits $C = \{2, \dots, 512\} \times \{0.6, \dots, 2.0\}$, in particular, we also limit $\rand$ over $\bsize$ to values from $8$ to $64$ with steps of $8$, and $\Delta$ to values between $0.8$ to $1.1$ with steps of $0.1$. The hyper-parameters escape from these values using $\mutate$ with $\alpha = 1.5, \beta=1.07, p=0.8$. It fixes $\gamma=16$ and $\delta=8$, $\bsize=3$, and $\Delta$ for hyper-parameter optimization with BS. The $\maxvisits$ parameter of Alg.~\ref{alg/bs} is bounded using previous optimizations procedures, i.e., the number of evaluations needed by the previous best configuration is doubled and defines $\maxvisits$ until a new optimization procedure is run.

\subsection{Pareto optimal $\fitness$ functions and the minimum quality $\fitness$ function}
A central component of our approach is the $\fitness$ function that will be used to navigate the $(C, \N)$ graph. As commented, we have identified two desirable behaviors: configurations that optimize both search speed and quality, having the best of both scores without compromising the other, and also configurations that achieve a minimum quality.

The best configuration that takes care of these two objectives and that cannot improve in one without decay in the other is called Pareto optimal. This optimal point depends on the precise characteristics of the workload, that is, the dataset, measure distance, and the query set used to probe.

For this matter, it is important to remark that these functions work under a determined context:
\begin{itemize}
    \item They work on a non-empty index using the a given set of parameters. We should measure costs and performances to create a metric.
    \item We work under the reasonable assumption that the test set (queries) will follow the dataset's distribution. Therefore, we use a small subset $Q$ of objects from the current indexed ones to compute the objective function, i.e., each $\fitness$ evaluation solves $|Q|$ $k$-nearest neighbors queries.
    \item The number of retrieved neighbors for $Q$ is the same as the test set.
    \item Recall-based objective functions compute a gold-standard result set on $Q$ with an exhaustive evaluation.
    \item We fix $\maxvisits$ to $O(\lceil \log^3{n} \rceil)$ for stopping any similarity search procedure that surpasses this cost.\footnote{Here $n$ is the current number of elements already indexed.} The cubic logarithmic cost is suggested in \cite{MPLK12} as the expected cost of the NSW.
\end{itemize}

The design of fitness functions is quite flexible since BS optimization algorithms do not need a gradient computation. In particular, we propose three fitness functions: 
\paragraph{\textsf{Pareto-recall}} Bi-objective optimization that searches for configurations that optimize search speed and recall. It is defined as:
$$\ParetoRecall((\bsize, \Delta)) = \left(\frac{\mathsf{visits}}{\maxvisits} \right)^2 + \left(1 - \recall\right)^2,$$
where \textsf{visits} is the average number of distance evaluations performed to solve a set of queries using a search graph with $\bsize$ and $\Delta$ as hyper-parameters. 
\paragraph{\textsf{Pareto-radius}} This is a bi-objective optimization that searches for configurations that improve search speed and minimizes the average radius of the result set. The optimization procedure is faster than Pareto-recall since it does not require computing a gold standard.
It is defined as: 
$$\ParetoRadius((\bsize, \Delta)) = \left(\frac{\mathsf{visits}}{\maxvisits}\right) + \left(\frac{\textsf{avg.  radius}}{\textsf{max. radius}}\right),$$
where $\mathsf{avg.}$ $\mathsf{radius}$ is the covering radius of the query set $Q$. The maximum covering radius must be estimated to avoid the computation of a gold standard, in particular, we use the first $\ParetoRadius$ evaluation in an optimization procedure to estimate it.
\paragraph{\textsf{Min-recall}} Optimizes an index to achieve a minimum quality, \textsf{min-recall}, in a training gold-standard; once the restriction is achieved, it looks for the faster configuration. This $\fitness$ function is defined as follows:
$$\MinRecall((\bsize, \Delta)) = \begin{cases}
    \recall < \textsf{min-recall}    & \rightarrow 3 - 2 \recall \\
    \recall \geq \textsf{min-recall} & \rightarrow \left(\frac{\mathsf{visits}}{\maxvisits} \right)
\end{cases} $$

The \textsf{min-recall} is a specified parameter while $\recall$ and $\mathsf{visits}$ are computed using a search graph with the given configuration.

Please note that our bi-objective functions require that both objectives remain on a similar scale.

Finally, please recall that $k$ nearest neighbor searches are represented as a minimization procedure using $\fitness_q(u) = \dist(q, u)$ for some $q \in U$ and some $u \in S \subseteq U$ while the fitness function for evaluating hyper-parameters use a working index and a small set of queries.
\FloatBarrier


\section{Experimental results}
\label{sec/experimental-results}

%
%


In this section, we describe the results of the experiments using Auto-tuning on the BS (ABS). We start on the build time and the memory needed for storing the index, and then we describe the performance of different combinations of fitness functions on each of the datasets,

In Table \ref{tab/autotuned-construction}, we show, for every dataset and each of the three different fitness functions, the amount of memory the index needs and the time are taken to build each index. Note that $\ParetoRadius$ produces the smaller indices since the covering radius of each query is minimized, which can promote retrieving fewer items than needed. Please recall that our construction algorithm is based on searches. This size reduction is hardly noticed because the dataset dominates the required memory. Regarding the building time, we can see a significant impact of the $\ParetoRadius$; for example, on DeepImage, the building time is $67\%$ that of $\MinRecall$ and only $35\%$ compared to $\ParetoRecall$. These gains can be seen over the other datasets. The $\MinRecall$ improves those $\ParetoRecall$'s construction costs on DeepImage, Lastfm, and WIT.

\begin{table}[!ht]
\caption{ABS construction costs using 64 threads for different fitness functions; in particular, we set $r=0.9$ for \textsf{Min-Recall}. These experiments fixes $b=1.2$.}
\label{tab/autotuned-construction}
\centering

\resizebox{\textwidth}{!}{
\begin{tabular}{r rr rr rr}
\toprule
    &  \multicolumn{3}{c}{\textsf{memory}} & \multicolumn{3}{c}{\textsf{build time}} \\ \cmidrule(lr){2-4} \cmidrule(lr){5-7}
   &\bf\textsf{Min-Recall}&\bf\textsf{Pareto-recall}&\bf\textsf{Pareto-radius}
   &\bf\textsf{Min-Recall}&\bf\textsf{Pareto-recall}&\bf\textsf{Pareto-radius}\\ \midrule
DeepImage-10M  &    5.0G  &   5.1G &    4.9G  &    344s &  573s & 204s \\
      GIST-1M  &    3.7G  &   3.7G &    3.7G  &    399s &  177s &  37s \\
   Glove-1.1M  &    651M  &   629M &    572M  &     85s &   50s &  12s \\
   Glove-400K  &    194M  &   191M &    181M  &     11s &    8s &   4s \\
  Lastfm-300K  &     92M  &    92M &     90M  &      5s &    8s &   2s \\
   Twitter-2M  &    2.8G  &   2.8G &    2.8G  &    180s &  174s &  52s \\
     WIT-300K  &    632M  &   633M &    630M  &     11s &   17s &   7s \\
\bottomrule
\end{tabular}
}
\end{table}

\begin{figure*}
\centering
{\includegraphics[width=\textwidth,trim={0cm 23cm 0 0},clip]{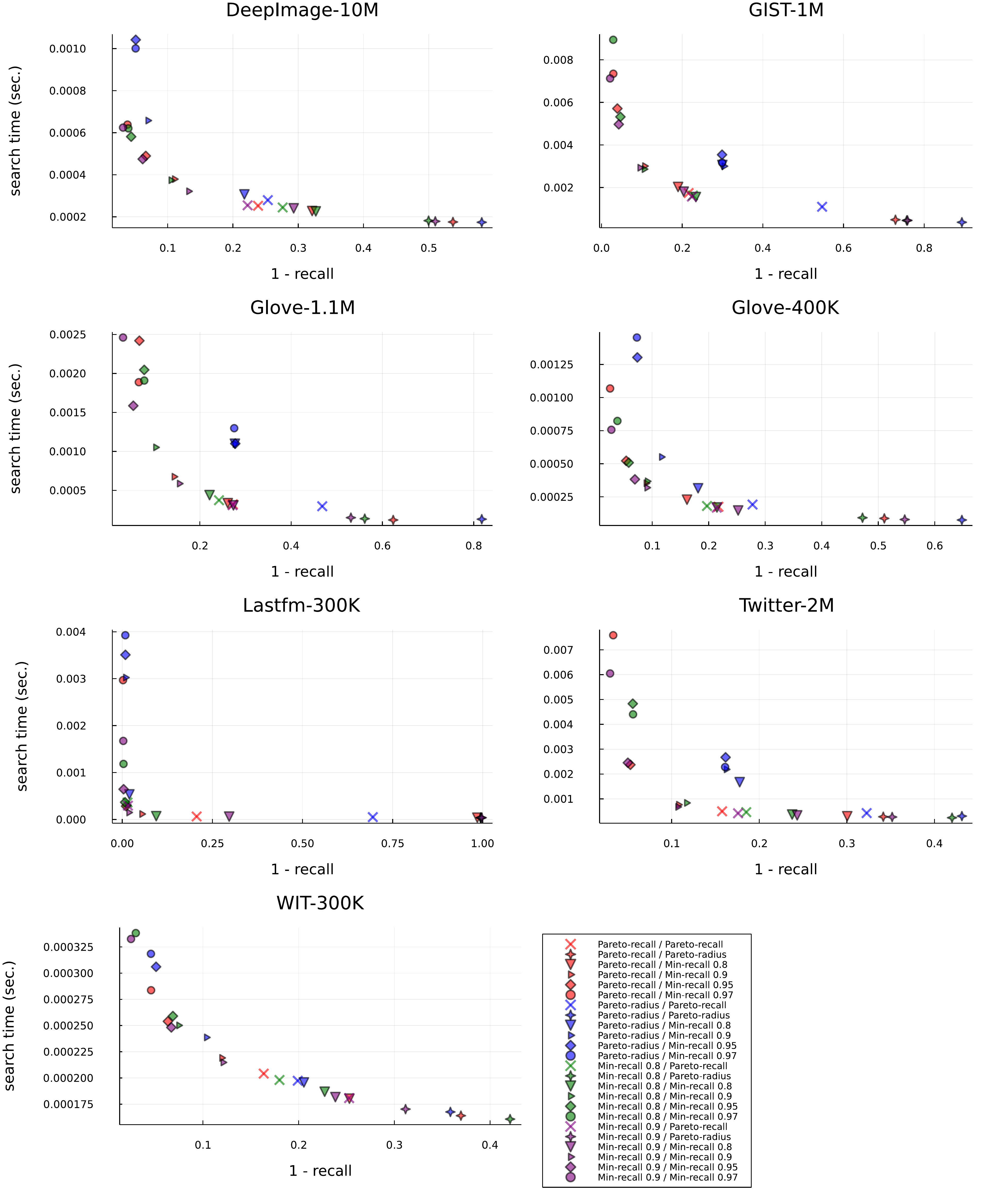}}
\caption{Search performance for our benchmarks using different kinds of auto-tuning parameters at both construction and searching stages.
    \label{fig/search-autotuned}
}
\end{figure*}

We now present the results for ABS on each of the datasets except BigANN. We used combinations of the $\ParetoRecall$, $\ParetoRadius$, and $\MinRecall$ fitness functions for building and searching. The recalls used on $\MinRecall$ were $ 0.8, 0.9, 0.95$ and $0.97$. Figure \ref{fig/search-autotuned} shows graphs where the $x$-axis is the error in the form of $1-$recall, and in the $y$-axis is the average search time in seconds. Please note that the graphs have different scales. The points closer to the origin represent better configurations of the BS. Let us start with the $\ParetoRecall$ fitness (red cross in the figure) on the construction and the search. Please remember that $\ParetoRecall$ has two objectives, i.e., improves search speed and recall jointly. Figure \ref{fig/search-autotuned} uses a color scheme to show the different constructions and markers indicating the kind of optimization made previously to solve the test queries. The red colors indicate a $\ParetoRecall$ on the construction step, and the cross indicates a $\ParetoRecall$ on the searching side. Blue color and stars represent $\ParetoRadius$. Different colors and markers represent each radius for $\MinRecall$. Now we can start to collect the findings from Figure \ref{fig/search-autotuned}. 

Let us start with the crosses. Recalls of the four crosses on each plot have relatively low recall and fast queries; for instance, in DeepImage and Glove-400k, their recall scores range between $0.7$ and $0.8$, but they also have shorter search times. On almost all the datasets, the blue cross got significantly lower recalls than the others, often without a gain in speed. 

The stars have $\ParetoRadius$ on the search; they are pretty fast but with the lowest recall. These combinations are not recommended for retrieving high-quality neighbors.

Please note that $\MinRecall$ of $0.8$  on the search optimization (i.e., the upside-down triangle) is pretty close to what we required. That is, these marks are close to $0.8$ (0.2 to be precise, since we are plotting $1 - \recall$) for the DeepImage and Glove-1.1M. The rest of the marks change the $\MinRecall$ to $0.9, 0.95,$ and $0.97$. We found that recall scores for the test query set are pretty similar to those expected ones. Note that optimization procedures (construction and previous to search) never saw the test set. These performances show the convenience of our automatic hyper-parameter tuning. 

Nonetheless, we must comment on the Lastfm-300k because it shows a different behavior than the rest of the datasets where we can see a smooth Pareto front. We found a sharp curve, and therefore, it is hard to have suitable configurations for Pareto since the area where a configuration excels at both is small. The optimization fails since it has to generalize from a sample in the training partition to the complete test set. Nonetheless, please look at the red, green, and purple right-facing triangles. These three optimizations appear very close to the origin, not just on Lastfm but on all our benchmarks.

In Table \ref{tab/neighborhood-statistics} we present some statistics for the neighborhoods created after using the auto-tuning BS. We show the mean number of neighbors and all the five data from the quartile statistics for each dataset. Please note that some of them have long tail distribution, notoriously GIST, Lastfm, and BigANN-100M.
\begin{table}[!ht]
    \centering
\resizebox{0.8\textwidth}{!}{
\begin{tabular}{rrrr rrrr}
\toprule
              &          & \multicolumn{6}{c}{\bf neighborhood sizes $|\N|$} \\ \cmidrule(lr){3-8}
\bf  name     & \bf kind &\bf  mean & \bf min. &\bf 1st q. &\bf median &\bf 3rd q. &\bf max. \\ \midrule
    BigANN-100M &      ABS &               36.8 &           1 &         20 &             30 &         45 &       1,013\\
      BigANN-1M &      ABS &               26.1 &           1 &         15 &             22 &         32 &         340\\
  DeepImage-10M &      ABS &               25.6 &           1 &         13 &             20 &         32 &         398\\
        GIST-1M &      ABS &                 11 &           1 &          4 &              7 &         13 &       1,631\\
     Glove-1.1M &      ABS &               31.3 &           1 &         17 &             25 &         38 &         666\\
     Glove-400K &      ABS &               18.7 &           1 &          9 &             14 &         24 &         555\\
    Lastfm-300K &      ABS &                  9 &           1 &          3 &              6 &         11 &       2,335\\
     Twitter-2M &      ABS &                 20 &           1 &         10 &             15 &         24 &         593\\
       WIT-300K &      ABS &               16.3 &           1 &          8 &             13 &         20 &         377\\
\bottomrule
\end{tabular}
}
\caption{Neighborhood statistics for the Beam Search index with automatic tuning index using a logarithmic neighborhood, we fixed $b=1.2$ and the SAT as the reducer.}
\label{tab/neighborhood-statistics}
\end{table}

%
%
\subsection{Comparison with other indices}
\label{sec/experimental-comparison-other-indices}

In this section, we show a comparison of our ABS with other state-of-the-art indexes in the categories of build time, memory, search speed, and recall. We tested the NNdescent implementation from PyNNDescent, the HNSW from the FAISS library, and SCANN from its official git repository. Each of these indexes was configured following the recommendations from their developers. In particular, for the NNdescent we fixed $divprob=1.0$ and $prun=1.5$, we vary the number of neighbors $N$. In the case of the HNSW, we used fixed values of $M=32$ and $efC=500$ and vary the $efS$ parameter using the values $16, 32, 64,$ and $128$. Finally, for the SCANN, we used $2\sqrt{n}$ as the number of leaves, the entire dataset for learning the codebook, and we varied the number of leaves for searching, i.e., 10, 50, 100, 300, 1000.

The constructions for all the indexes used 64-threads, but the searches were made using a single thread. The ABS was constructed with \textsf{Pareto-recall} and then optimized for searching using \textsf{Min-recall}; a single index per dataset was created. We use a single index for other datasets with their search parameters vary, with the exception of NNdescent that we need to create several indexes.

Table \ref{tab/autotuned-low-memory} shows the building time (in seconds) and the memory (in MB) of our ABS with NNdescent, HNSW, and SCANN over each of the datasets. The results for the NNdescent are the average of four different configurations where we varied the number of neighbors (marked with an asterisk). We will start with the BigANN dataset, which has not been studied above. This dataset is large enough to make NNdescent unable to create its index, despite our machine having relatively high resources. All the indexes have a similar building time and memory, with the SCANN being the more appealing option. The following dataset on Table \ref{tab/autotuned-low-memory} is again the BigANN but with one million elements. Note the building time first; the ABS takes only 16 seconds, the SCANN takes almost double that, the HNSW takes 91, and the NNdescent needs 461 seconds. This behavior repeats in the rest of the datasets. The ABS has the best building times, closely followed by the SCANN. The NNdescent is left behind in these benchmarks. The memory needed to store the index has a similar tendency; the ABS and SCANN are pretty close on all the datasets.

\begin{table}[!ht]
\caption{Performance comparison for our ABS and state-of-the-art alternatives on build time (in seconds) and memory (in MB).
}
\label{tab/autotuned-low-memory}
\begin{minipage}[t][][t]{0.33\textwidth}
\resizebox{\textwidth}{!}{
\begin{tabular}{rrr}
\toprule
       & build  &        \\
 method & time  & memory \\
\midrule \multicolumn{3}{c}{BigANN-100M } \\ \midrule
    ABS     &   12,770  &  64,395 \\
    NNdescent & -       & - \\
    HNSW    &   14,441  &  62,587 \\
    SCANN   &   11,024  &  55,333 \\
\midrule \multicolumn{3}{c}{BigANN-1M } \\ \midrule
    ABS      &     16  &      604 \\
   NNdescent &   *461  &   *1,280 \\
        HNSW &     91  &     748  \\
SCANN        &     30  &     555  \\
\midrule \multicolumn{3}{c}{DeepImage-10M } \\ \midrule
  ABS            &    172  &    4,786 \\
  NNdescent      & *4,476 & *10,151   \\
      HNSW    &      859 &   6,251 \\
    SCANN     &      300 &   4,159 \\
\bottomrule
\end{tabular}
}
\end{minipage}~\begin{minipage}[b][][b]{0.318\textwidth}
\resizebox{\textwidth}{!}{
\begin{tabular}{rrr}
\toprule
       & build  &        \\
 method & time  & memory \\

\midrule \multicolumn{3}{c}{GIST-1M } \\ \midrule
 ABS            &      202  &    3,720 \\
   NNdescent    &     *548  &   *8,249 \\
       HNSW     &      734  &   3,922 \\
    SCANN       &      231  &   4,138 \\

\midrule \multicolumn{3}{c}{Glove-1.1M } \\ \midrule
    ABS         &   35 &      611 \\
   NNdescent    & *635 &    *1,305 \\
       HNSW     &  161 &     759  \\
    SCANN       &   21 &     514  \\

\midrule \multicolumn{3}{c}{Glove-400K } \\ \midrule
    ABS        &    6 &  183 \\
    NNdescent  & *234 & *398 \\
      HNSW   &    24  &  250 \\
    SCANN    &   8.9  &  170 \\
\bottomrule
\end{tabular}
}
\end{minipage}~\begin{minipage}[b][][b]{0.33\textwidth}
\resizebox{\textwidth}{!}{
\begin{tabular}{rrr}
\toprule
       & build  &        \\
 method & time  & memory \\
\midrule \multicolumn{3}{c}{Lastfm-300K } \\ \midrule
    ABS       &    3 &   88 \\
    NNdescent & *222 & *191 \\
       HNSW   &  4.7 &  148 \\
   SCANN      & 3.7  &   83 \\

\midrule \multicolumn{3}{c}{Twitter-2M } \\ \midrule
    ABS         &   118 &    2,797 \\
    NNdescent   & *1,363 &   *6,021 \\
     HNSW       &  630  &    3,176 \\
    SCANN       &  123  &    2,927  \\
\midrule \multicolumn{3}{c}{WIT-300K } \\ \midrule
    ABS        &    20 &   627 \\
    NNdescent  &  *193 & *1,366 \\
        HNSW   &   53  &   682 \\
    SCANN      &      33  &    683  \\ \bottomrule
\end{tabular}
}
\end{minipage}
\end{table}

\begin{figure*}
    \centering
    \includegraphics[width=0.8\textwidth]{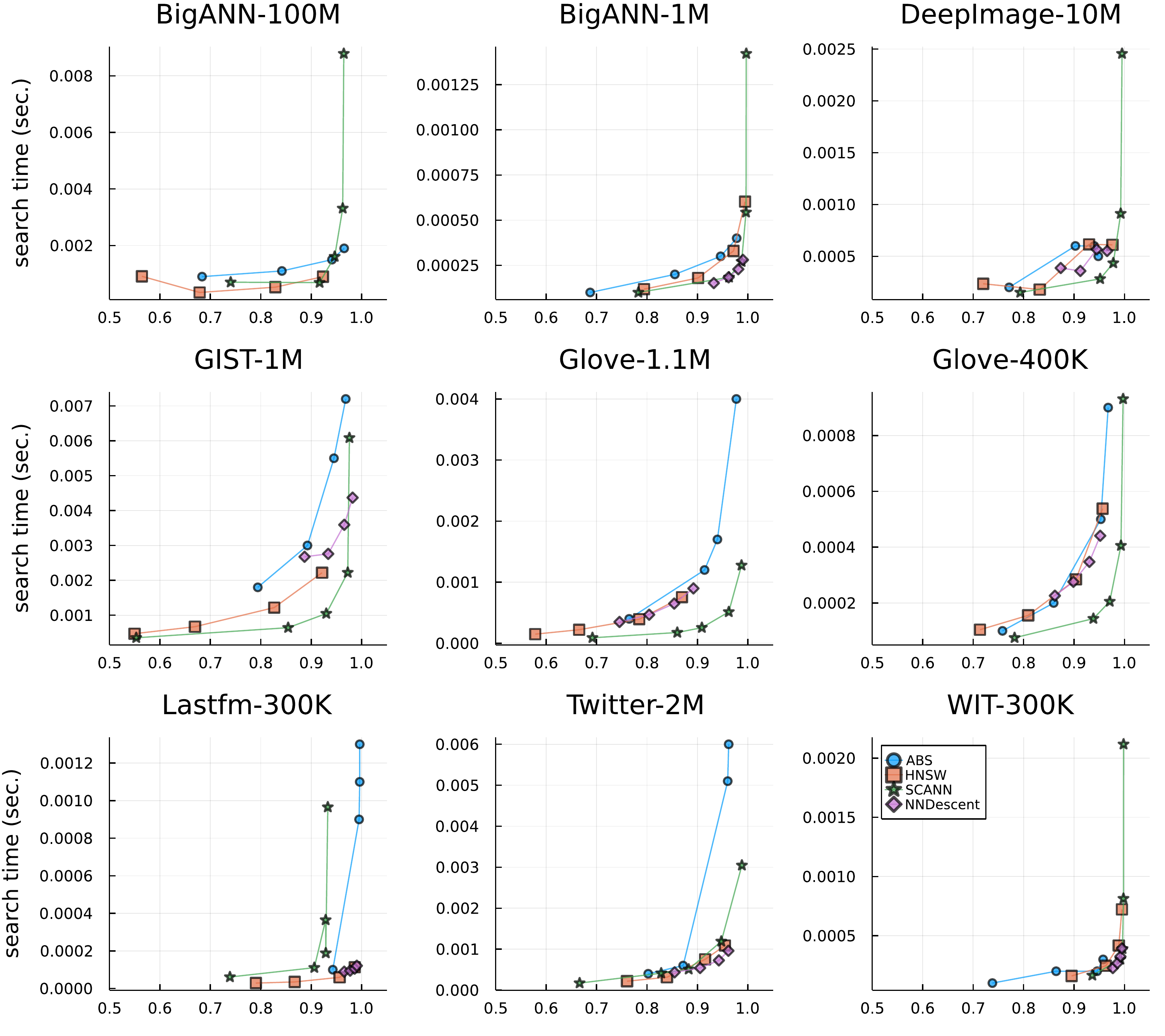}
    \caption{Search time and recall comparison between ABS and other state-of-the-art methods on different benchmarks.}
    \label{fig/sota-comp}
\end{figure*}

Figure \ref{fig/sota-comp} presents a comparison regarding search speed and recall for each benchmark. On the $x$-axis, we have the recall, and on the $y$-axis, the search time (seconds). Better performances are close to the bottom-right corner.

The first dataset appearing is the BigANN with 100 million elements. Here, the HNSW is quicker but has the worst recalls. The ABS and the SCANN have similar recalls at comparable speed. Note, for example, that the ABS can reach like 0.97 on recall with faster searches than the SCANN. This dataset helps to show the strengths of the ABS; first, its usage is pretty simple, the hyper-parameters are selected automatically for the desired recall; also, it is very competent on the building step. It also has the advantage of working with any distance function. Indexes like SCANN work better for inner product distance.

For the BigANN with one million elements, the NNdescent has the quickest searches at high-recall scores, followed by SCANN, the ABS get comparable speeds than that achieved with HNSW.
On our next benchmark, DeepImage, the SCANN and ABS got the best recall scores. If we are looking for an index that delivers more than $0.96$ recall, the best choice is SCANN, however it requires to be manually tuned for this. Regarding search speed, the results of ABS, HNSW and NNdescent are similiar.

For the GIST dataset, SCANN achieves faster searches again. The HNSW fails to achieve a recall of at least $0.95$ but also achieves fast searches. The recalls of ABS and SCANN are close. Regarding speed, the ABS is the slower one in GIST (2 times slower than the faster approach). This slow down can be explained by the power-law distribution found in the neighborhood's size distribution, see Table~\ref{tab/neighborhood-statistics}.

The results for the Glove dataset are next. First, with 1.1 million embeddings. Here the HNSW and the NNdescent got recalls below the $0.9$ mark. The Glove with 400K vectors appears to the side, and all indexes achieve better recall scores. The SCANN got faster queries than other indexes.
The search speed of ABS performance is similar to HNSW and NNDescent for Glove datasets; however, it achieves these recall scores since these were asked and automatically tuned.

The Lastfm with 300K embedding vectors has the smallest dimension among our benchmarks, i.e., $65$. The ABS, NNdescent, and HNSW all got excellent recall scores on this tiny dataset. Surprisingly, SCANN achieves the lower recall scores. Lastfm is hard to optimize automatically with Pareto, but we also used \textsf{Min-recall} before searching.
The penultimate plot in Figure \ref{fig/sota-comp} is that of the Twitter-2M benchmark. All indexes perform similarly for recalls below $0.9$, yet the NNdescent has slighter faster searches. Only ABS and SCANN achieve recall scores higher than 0.96, but high recall setups require significantly more computing power.

Finally, all indices achieve similar performance for the WIT benchmark. All got high-quality recalls, with the NNdescent achieving faster searches. Note how the searches of ABS with high-recall scores are very close in time with the NNdescent and HNSW for searches with comparable recall. Please note that ABS show lower recall scores since they are part of our experimental setup; that is, we ask for them.

\subsection{Optimized parameters}
\label{sec/optimized-parameters}

The ABS optimizes $\bsize$ and $\Delta$ parameters during the construction and previous to searching. Table \ref{tab/opt-abs} shows the final $\bsize$ and $\Delta$ used for searching. These values were found by minimizing $\MinRecall$ with $r=0.8$, $0.9, 0.95$ and $0.97$. The table also shows several performance measures that characterize our method:  the average number of visits (distance computations) for the searches, the time spent by the $\MinRecall$ optimization, the queries per second, and the train and test recall scores.
Please look the table's rows related with BigANN datasets; the $\bsize$ and $\Delta$ are similar despite the 100 times difference in size. Glove and Twitter benchmarks are all word embeddings with similar sizes; they all have differences in the hyperparameters. Please note that these values are independent of the dataset size and dependent on the dataset distribution and complexity. This tuning complexity is tackled automatically with our approach.

On the other hand, note that the average number of visits increases with $r$ (the queries per second decrease since they are inversely proportional to visits). In general, we observe that optimization time is short and proportional to $r$ except for BigANN-100M, where the large size could explain this divergence.

Table \ref{tab/opt-abs} also shows that the recall test closely follows the value of the recall train, and it is also pretty similar to the optimized $r$. The optimization for $r=0.8$ gives the bigger differences between the train and test recalls. This difference might indicate overfitting over the training data. Please remember that Pareto optimal points are around this value; therefore, slight movement in the configuration space may produce significant performance changes in an index. Again, the Lastfm benchmark is distinctive since train recall scores are similar to $r$, but this is not reproduced in the test. In general, it achieves extreme bad or good results. 

\begin{table}
\caption{Performance and characteristics of our ABS with several configurations and benchmarks.}
\label{tab/opt-abs}
\centering
\resizebox{0.7\textwidth}{!}{
\begin{tabular}{lrlr lrll}
\toprule
     &       &          & avg.        & opt.      &         & train        & test        \\
kind & bsize & $\Delta$ &      visits &      time & q/s     &       recall &      recall \\

\midrule\multicolumn{8}{c}{BigANN-100M} \\ \midrule
  ABS   $r=0.8$ &       8 &   0.9346 &       1,163 &    43 &      995 &      0.8106 &   0.6674\\
  ABS   $r=0.9$ &      12 &      1.1 &       2,199 &  32.1 &      737 &      0.9044 &   0.8633\\
  ABS  $r=0.95$ &      45 &   1.1529 &       4,718 &    25 &      481 &      0.9583 &   0.9534\\
  ABS  $r=0.97$ &      51 &   1.2027 &       6,677 &    25 &      408 &      0.9725 &   0.9700\\ 
\midrule \multicolumn{8}{c}{BigANN-1M} \\ \midrule
  ABS   $r=0.8$ &       9 &   0.8486 &         596 &   0.3 &    9,331 &      0.8026 &   0.6924\\
  ABS   $r=0.9$ &      11 &   1.0656 &       1,132 &   0.3 &    5,576 &      0.9001 &   0.8843\\
  ABS  $r=0.95$ &      23 &   1.0772 &       1,382 &   0.4 &    4,290 &      0.9508 &   0.9311\\
  ABS  $r=0.97$ &      56 &   1.1206 &       2,356 &   0.4 &    2,629 &      0.9721 &   0.9727\\ 
\midrule \multicolumn{8}{c}{DeepImage-10M} \\ \midrule
  ABS   $r=0.8$ &      11 &   0.9509 &       1,058 &   0.9 &    5,008 &      0.8149 &   0.7446\\
  ABS   $r=0.9$ &      30 &   1.0512 &       2,118 &   1.2 &    2,856 &       0.902 &    0.8990\\
  ABS  $r=0.95$ &      41 &      1.1 &       2,494 &   1.1 &    2,198 &      0.9541 &   0.9346\\
  ABS  $r=0.97$ &      73 &   1.2594 &       7,842 &   1.5 &      997 &      0.9706 &   0.9837\\ 
\midrule \multicolumn{8}{c}{GIST-1M} \\ \midrule
  ABS   $r=0.8$ &      40 &        1 &       2,375 &   2.2 &      525 &      0.8035 &   0.7518\\
  ABS   $r=0.9$ &      56 &      1.1 &       6,192 &   5.1 &      260 &      0.9001 &   0.9041\\
  ABS  $r=0.95$ &      44 &    1.177 &       9,878 &   4.7 &      155 &      0.9536 &    0.9460\\
  ABS  $r=0.97$ &     104 &    1.177 &      11,473 &     9 &      127 &      0.9702 &   0.9649\\ 
\midrule \multicolumn{8}{c}{Glove-1.1M} \\ \midrule
  ABS   $r=0.8$ &      26 &   1.0272 &       2,678 &   0.6 &    2,621 &      0.8007 &   0.7752\\
  ABS   $r=0.9$ &      28 &      1.1 &       3,925 &   0.4 &    1,574 &      0.9034 &   0.8537\\
  ABS  $r=0.95$ &     114 &    1.177 &      15,846 &   1.3 &      485 &      0.9512 &   0.9516\\
  ABS  $r=0.97$ &     356 &    1.177 &      23,791 &   2.9 &      290 &      0.9692 &   0.9693\\ 
 
\midrule \multicolumn{8}{c}{Glove-400K} \\ \midrule
  ABS   $r=0.8$ &      13 &   0.9798 &         854 &   0.1 &    7,716 &      0.8191 &   0.7574\\
  ABS   $r=0.9$ &      16 &      1.1 &       1,946 &   0.2 &    4,110 &      0.9091 &   0.8747\\
  ABS  $r=0.95$ &      84 &      1.1 &       3,170 &   0.4 &    2,138 &      0.9503 &   0.9355\\
  ABS  $r=0.97$ &     102 &   1.1847 &       6,085 &   0.7 &    1,309 &      0.9711 &   0.9658\\ 
\midrule \multicolumn{8}{c}{Lastfm-300K} \\ \midrule
  ABS   $r=0.8$ &      19 &   0.9813 &         529 &   0.1 &   37,360 &       0.803 &   0.3688\\
  ABS   $r=0.9$ &      16 &   1.1096 &         609 &   0.1 &    1,335 &      0.9006 &   0.9937\\
  ABS  $r=0.95$ &      42 &   1.2165 &       1,240 &   0.2 &      447 &      0.9598 &   0.9969\\
  ABS  $r=0.97$ &      90 &   1.3183 &       2,027 &   0.6 &      143 &      0.9702 &   0.9979\\ 
\midrule \multicolumn{8}{c}{Twitter-2M} \\ \midrule
  ABS   $r=0.8$ &      10 &   0.9971 &       1,021 &     1 &    3,162 &      0.8011 &    0.7730\\
  ABS   $r=0.9$ &      14 &   1.0273 &       1,279 &     1 &    2,405 &       0.901 &   0.8276\\
  ABS  $r=0.95$ &      84 &   1.2182 &       6,925 &   1.5 &      359 &      0.9403 &   0.9505\\
  ABS  $r=0.97$ &     190 &   1.3475 &      20,450 &   3.3 &       97 &      0.9697 &   0.9766\\ 
\midrule \multicolumn{8}{c}{WIT-300K} \\ \midrule
  ABS   $r=0.8$ &       5 &   0.8569 &         373 &   0.3 &    8,571 &      0.8054 &   0.7386\\
  ABS   $r=0.9$ &      11 &   0.9548 &         514 &   0.3 &    6,242 &      0.9124 &    0.8930\\
  ABS  $r=0.95$ &      21 &    1.046 &         741 &   0.3 &    4,162 &      0.9678 &   0.9621\\
  ABS  $r=0.97$ &      23 &    1.035 &         801 &   0.6 &    4,064 &      0.9716 &    0.9620 \\ \bottomrule
\end{tabular}
}
\end{table}

\section{Conclusions}
\label{sec/conclusions}

This manuscript introduces a graph-based similarity search index that efficiently uses the Beam Search algorithm to solve $k$ nearest neighbor queries. We modified the existing algorithms and structures from our previous work. We added a new expansion hyperparameter called $\Delta$ that allowed trading search speed and accuracy, and introduced an automatic hyperparameter optimization based on the Beam Search. This optimization metaheuristic allows fast constructions with an expected performance of the index. We provide two kinds of optimizations, bi-objective Pareto optimality setups for search speed and quality; we also produce setups for achieving fast queries with a minimum expected quality. We aim to simplify similarity search indexes while obtaining fast and accurate indexes. We provide an opensource implementation under the MIT license for the Julia language.

We provide extensive experimental evidence that characterizes and compares our contribution in various benchmarks corresponding to vision, language, and vision \& language tasks. More detailed, we compare our approach with three state-of-the-art methods on nine real-world datasets achieving competitive results in construction cost, memory, search time, and recall.

It is worth mentioning that our approach achieves convenient tradeoffs in almost all our benchmarks, but not in one, Lastfm. We had observed a sharped curve in the configuration space that makes it hard to tune automatically. Another issue we are aware of is the overfitting effect found in several datasets. While this issue is expected since we use objects in the train set to adjust hyperparameters, the approach needs more work on regularization methods. Finally, we also observed skewed distributions of neighborhood sizes, some of them following a power law. This distribution limits the search speed in two ways; firstly, low degree nodes have low routing capabilities while high degree nodes will need too much time to be explored. More research is needed to reduce the skewness of this distribution and improve search times.

\bibliographystyle{elsarticle-num}
\small
\bibliography{biblio}

\end{document}